\documentclass[nofootinbib,aps,superscriptaddress,preprint,floatfix]{revtex4}
\usepackage{epsfig}
\usepackage{amsmath}
\usepackage{setspace}
\usepackage{color}
\usepackage{subfigure}
\usepackage{graphicx}

\usepackage{epstopdf}
\usepackage{pstricks}
\def\beq{\begin{equation}}
\def\eeq{\end{equation}}
\def\bea{\begin{eqnarray}}
\def\eea{\end{eqnarray}}
\def\beqa{\begin{eqnarray}}
\def\eeqa{\end{eqnarray}}
\def\dslash{\not\negmedspace}

\begin{document}
\vspace{3.0cm}
\preprint{\vbox {
\hbox{WSU--HEP--1202} 
%\hbox{hep-ph/yymmddd}
}}

\vspace*{2cm}

\title{\boldmath Faking $B_s \to \mu^+\mu^-$}

\author{Y.G. Aditya \vspace{5pt}}
\email{ygaditya@wayne.edu}
\affiliation{Department of Physics and Astronomy\\[-6pt]
        Wayne State University, Detroit, MI 48201}

\author{K.J. Healey\vspace{5pt}}
\email{healey@wayne.edu}
\affiliation{Department of Physics and Astronomy\\[-6pt]
        Wayne State University, Detroit, MI 48201}

\author{Alexey A.\ Petrov\vspace{5pt}}
\email{apetrov@wayne.edu}
\affiliation{Department of Physics and Astronomy\\[-6pt]
        Wayne State University, Detroit, MI 48201}
\affiliation{Michigan Center for Theoretical Physics\\[-6pt]
University of Michigan, Ann Arbor, MI 48109\\[-6pt] $\phantom{}$ }

\begin{abstract}
Recent evidence from observation of the flavor-changing neutral current decay $B^0_s \rightarrow \mu^+ \mu^-$ by the LHCb collaboration $\mathcal{B}(B^0_s \rightarrow \mu^+ \mu^-) = (3.2^{+1.5}_{-1.2})\times 10^{-9}$ is consistent with the latest standard model prediction $\mathcal{B_{SM}}(B^0_s \rightarrow \mu^+ \mu^-) = (3.23\pm0.27)\times 10^{-9}$.
While new physics can still affect this decay amplitude, its contribution is certainly not the dominant one. 
We analyze branching ratios of the decays $B^0_s \rightarrow \mu^+ \mu^- X$, with $X = \gamma$ or $\nu\bar\nu$, which can mimic $B^0_s \rightarrow \mu^+ \mu^-$ on portions of the parameter space where $X$ is soft. We perform a model-independent standard model calculation of those processes incorporating heavy quark and chiral symmetries of QCD. We show that the considered contributions contribute to $B^0_s \rightarrow \mu^+ \mu^-$ at a sub-percent level.
\end{abstract}

\def\thepage{{}}
\maketitle
\def\thepage{\arabic{page}}

%%%%%%%%%%%%%%%%%%%%%%%%%%%%%%%%%%%
\section{Introduction}

The rare leptonic decay of the $B^0_s$ into a dimuon pair, $B^0_s \to \mu^+\mu^-$, is an example of a flavor-changing neutral current (FCNC) process. Studies of such decay processes not only play an important role in determining electroweak and strong interaction parameters of the standard model (SM) of particle physics, but also serve as sensitive probes of possible physics beyond the standard model~\cite{Golowich:2011cx}. While recent evidence for observation of $B^0_s \rightarrow \mu^+ \mu^-$ from LHC-b collaboration~\cite{:2012ct} , as well as an earlier result from CDF~\cite{Aaltonen:2011fi} preclude any spectacular new physics (NP) effect, there is still room for NP to influence this decay. It is then important to have a firm evaluation of $\mathcal{B}(B^0_s \rightarrow \mu^+ \mu^-)$ in the SM~ \cite{Buras:2012ru,Shigemitsu:2009jy} and a firm understanding that experimentally-observed branching ratio 
\begin{eqnarray}
\mathcal{B}_{LHCb}(B^0_s \rightarrow \mu^+ \mu^-) &\,=\,& (3.2^{+1.5}_{-1.2})\times 10^{-9}
\nonumber \\
\mathcal{B}_{CDF}(B^0_s \rightarrow \mu^+ \mu^-) &\,=\,&  (1.8^{+1.1}_{-0.9})\times 10^{-8}
\end{eqnarray}
actually corresponds to the $B^0_s \rightarrow \mu^+ \mu^-$ transition.

It is well known that the $B^0_s \rightarrow \mu^+ \mu^-$ decay is helicity suppressed in the SM by $m_\mu^2/m_{B_s}^2$ due to the left handed nature of weak interactions~\cite{Buras}. This effect arises from the necessary spin flip on the outgoing back-to-back lepton pair in order to conserve angular momentum since the initial state meson is spinless.  

This suppression is absent in $B^0_s$ decays where the muon pair is produced with one or more additional particles in the final state that can carry away a unit of angular momentum, such as $B^0_s \rightarrow \mu^+ \mu^-\gamma$ or $B^0_s \rightarrow \mu^+ \mu^-\nu_\mu\bar\nu_\mu$. This means that, in general, those processes could have sizable total branching ratios, comparable to that of $B^0_s \rightarrow \mu^+ \mu^-$, despite being suppressed by other small parameters (such as $\alpha_{EM}$ for $B^0_s \rightarrow \mu^+ \mu^-\gamma$)~\cite{Melikhov:2004mk}. If, in addition, the final state photon or $\nu\bar\nu$ is undetected, while the invariant mass of $\mu^+ \mu^-$ pair is close to $m_{B^0_s}$, then the experimentally-measured branching ratio would correspond to
\begin{equation}\label{ExpBr}
\mathcal{B}_{exp}(B^0_s \rightarrow \mu^+ \mu^-)  = \mathcal{B} (B^0_s \rightarrow \mu^+ \mu^-)  
\left[ 1+ \sum_X
 \frac{\mathcal{B} (B^0_s \rightarrow \mu^+ \mu^-X)|_{m(\mu^+\mu^-)\approx m_{B_s})} }{\mathcal{B} (B^0_s \rightarrow \mu^+ \mu^-) }
\right],
\end{equation}
where $X$ is an undetected particle or a group of particles. The contribution of $B^0_s \rightarrow \mu^+ \mu^-X$ would depend on how well $X$ could be detected in a particular experiment, as well as on whether $B^0_s \rightarrow \mu^+ \mu^-X$ has any kind of resonance enhancement that is not well modeled by background models chosen by a particular experiment in a given window of $m(\mu^+\mu^-)$, as well as the size of that window. For example, for $X=\gamma$, most current searches use di-lepton energy cuts that would correspond to an allowable soft photon of up to 60 MeV. 
For $B \to \ell\nu_\ell$ transition and $X=\gamma$ similar effects were discussed in~\cite{RadDecays,Burdman:1994ip,Chiladze:1998ny}, and for $X$ being light particles -- in \cite{Aditya:2012ay}. In the following we shall concentrate on the amplitudes that are non-vanishing in the $m_{\mu} \rightarrow 0$ limit.

%%%%%%%%%%%%%%%%%%%%%%%%%%%%%%%%%%%
\section{$B^0_s \to \mu^+\mu^-\gamma$ transition}

Due to higher backgrounds in hadron collider experiments soft photons emitted in $B^0_s \rightarrow \ell^+ \ell^-\gamma$ could be hard to detect, so this background could be quite important. This decays were previously analyzed in~\cite{Melikhov:2004mk}, where a form-factor model-dependent calculation was performed~\cite{Kruger:2002gf} . The analysis presented in \cite{Melikhov:2004mk} was mainly geared towards kinematical regimes where the photon is sufficiently hard to be detected; in fact, low-energy cut-offs were introduced on photon energies. We apply a model-independent approach that incorporates both heavy quark symmetry for hadrons containing a heavy quark with mass $m_Q >> \Lambda_{QCD}$, and chiral $SU(3)_L\times SU(3)_R$ symmetries in the $m_q\rightarrow 0$ limit~\cite{Wise:1992hn,Burdman:1992gh}. We organize our calculations in terms of an expansion in $1/m_b$ and examine the contribution of terms up to leading order in $\mathcal{O}(1/m_b)$.

\begin{figure}
\begin{center}
\subfigure[]{ \includegraphics[width=6.5cm]{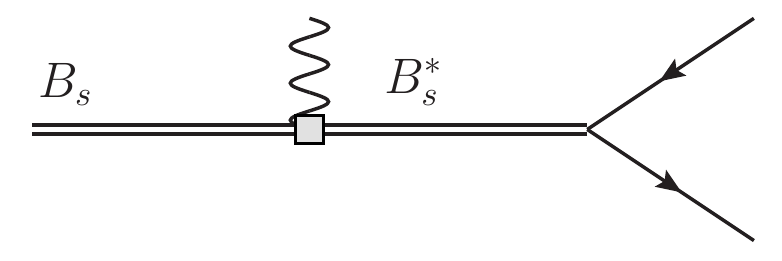}}
\subfigure[]{ \includegraphics[width=6.5cm]{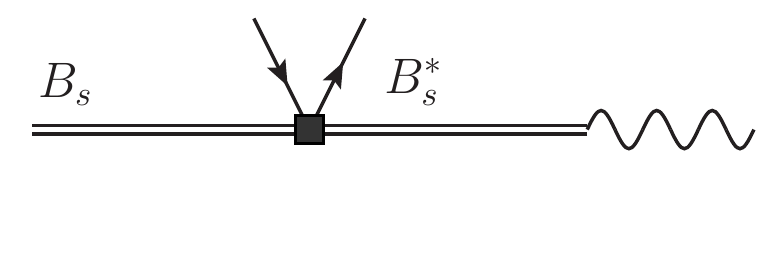}}
\subfigure[]{ \includegraphics[width=6.5cm]{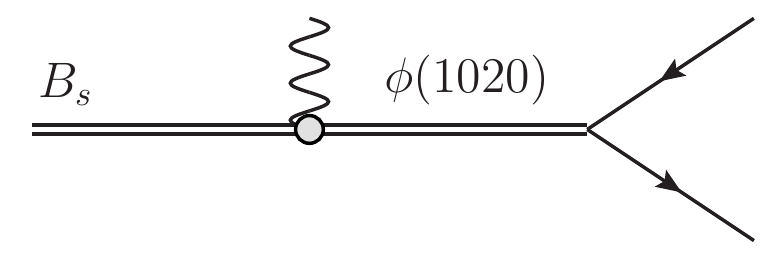}}
\subfigure[]{ \includegraphics[width=6.5cm]{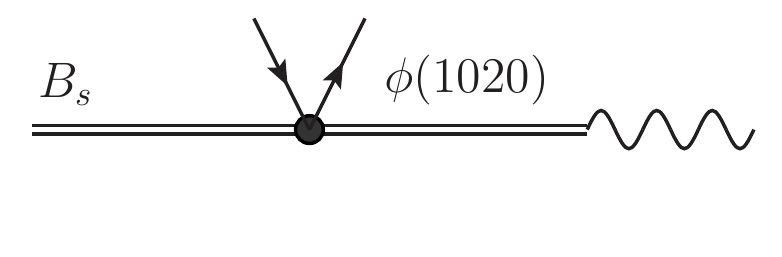}}
\end{center}
\caption{Resonant contributions to $B_s \rightarrow \mu^+ \mu^-\gamma$}
\label{FIG:pole} % caption for the whole figure
\end{figure}

Similarly to $B \to \ell\nu_\ell\gamma$~\cite{Burdman:1994ip}, the decay amplitude for $B_s \rightarrow \mu^+ \mu^-\gamma$ transition can be broken into two generic parts containing internal bremsstrahlung (IB) and structure dependent (SD) contributions. The bremsstrahlung contributions are still helicity suppressed, while the SD contribution contain the electromagnetic coupling $\alpha$ but are not suppressed by the lepton mass. Phenomenologically, the origin of that can be understood as follows. 
When the soft photon in emitted from the $B_s$ meson, heavy intermediate states including the $J^P = 1^{-}$ $B_s^*$ vector meson state become possible. This lifts helicity suppression since the lepton pair couples directly to the spin 1 meson. In the kinematic regime where the photon is soft, we expect that the significant contribution comes only from the vector $B_s^*$ resonance for reasons analogous to the $B^*$ pole dominance in $B\rightarrow\pi\ell\nu$ at near zero pionic recoil energies ~\cite{Isgur:1989qw}. This is because in the large $m_b$ limit the $B_s^*$ and $B_s$ become degenerate and the residual mass splitting is $m_{B^*}-m_B \sim \mathcal{O}(1/m_b)$~\cite{Georgi:1990um}. Therefore the excitation of the $B_s^*$ does not require much energy. There are two diagrams containing an intermediate $B_s^{*}$ as seen in Fig.~\ref{FIG:pole}. In the kinematic region of interest where $E_\gamma < 60 MeV$, Fig.~\ref{FIG:pole} (b) where the  intermediate $B_s^{*}$ decays to an on-shell soft photon is $(1/M_{B^0_s})$ suppressed and will be neglected. Similarly, a contribution of Fig.~\ref{FIG:pole} (d) is formally $(1/M_{B^0_s})$, so it will be neglected in what follows.

The calculation of soft photon effects should carefully deal with soft divergencies. Those are cancelled between
one-loop radiative corrections to $B_s \rightarrow \mu^+ \mu^-$ and $B_s \rightarrow \mu^+ \mu^-\gamma$.

\begin{figure}
\begin{center}
{\includegraphics[width=6.5cm]{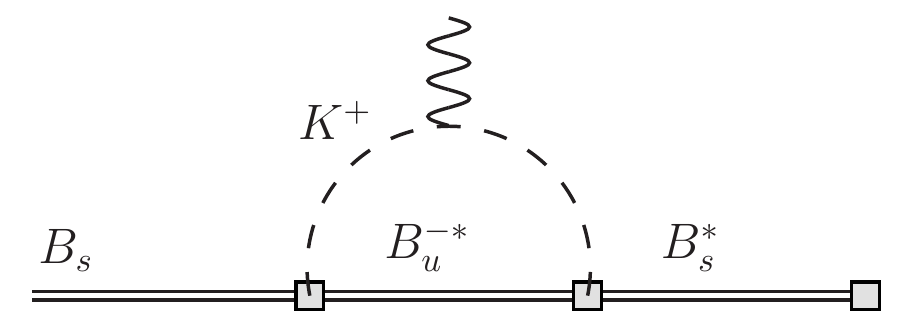}}
\end{center}
\caption{One loop corrections to $\mu$. The double lines denote the heavy mesons B and $B^*$ 
while the single line denotes the goldstone bosons}
\label{FIG:oneloop} % caption for the whole figure
\end{figure}

We employ Heavy Meson Chiral Perturbation Theory (HM$\chi$PT) to calculate Fig.~\ref{FIG:pole} (a). The heavy meson superfield 
$H_a$ contains both the $B^0_s$ and $B_s^{*0}$ bosons,
\beq
H_a = \frac{1 + \dslash v}{2} ( B_{a \mu}^* \gamma^{\mu} - B_a \gamma_5), \qquad\qquad \bar{H}_a = \gamma^0(H_a)^\dag\gamma^0,
\eeq
where the indices $a$ and $b$ reflect the light quark flavor indices. The light mesons are introduced through the matrix 
$Q^\xi = \frac{1}{2}(\xi^\dag Q\xi +\xi Q\xi^\dag)$ where the field $\xi = \exp(i\Pi/f)$ is defined in terms of a $3\times 3$ unitary matrix 
containing the octet of pseudo-Goldstone bosons
\beq
\Pi =\left( \begin{array}{ccc}  \frac{\pi^0}{\sqrt{2}}+\frac{\eta}{\sqrt{6}} & \pi^+ & K^+ \\ \pi^- & -\frac{\pi^0}{\sqrt{2}}+\frac{\eta}{\sqrt{6}} & K^0\\ K^- & \bar{K}^0 & -\sqrt{\frac{2}{3}}\eta \end{array}\right).
\eeq
To evaluate diagram Fig.~\ref{FIG:pole} (a) we need an amplitude for a $B\rightarrow B^*\gamma$ transition as
\beqa\label{BstarContr}
\mathcal{M}_{[B_s \rightarrow B_s^{*}\gamma_s \rightarrow \mu^+ \mu^- \gamma_s]} \,=\, 
\mathcal{M}^{\mu}_{B_s^{*} \rightarrow \mu^+ \mu^-}  
\times \frac{g_{\mu \alpha}}{M_{B_s^*}^2} \times
\mathcal{M}^{\alpha}_{B_s\rightarrow B_s^{*} \gamma} ,
\eeqa
The amplitude for $B\rightarrow B^*\gamma$ is conventionally parameterized as 
\beqa
\mathcal{M}_{B_s\rightarrow B_s^{*} \gamma}  = -i e \mu \eta^{*}_{\alpha} v_{\beta} k_\mu \epsilon_{\nu}^{*} \epsilon^{\mu \nu \alpha \beta},
\eeqa
where $k$ is the 4-momentum of the photon, $v$ the velocity of the decaying heavy meson, $\eta$ is the vector meson 
polarization, and $\epsilon$ is the photon polarization.  The strength of the transition is described by the magnetic 
moment, $\mu$, which receives contributions from the photon coupling to both heavy and the light quark components
of the electromagnetic current~\cite{Amundson:1992yp},
 \beq
\mu = \mu_{b} + \mu_{\ell}.
\eeq
The bottom quark contribution is fixed by heavy quark symmetry to be $\mu_b =  Q_b/m_b = -1/(3m_b)$, while the light quark
contribution can be computed, to one loop, in HM$\chi$PT. The relevant effective Lagrangian 
is \cite{Amundson:1992yp,Stewart:1998ke}
\beq
\mathcal{L}_\beta = \frac{\beta e}{4} \mbox{Tr}(\bar{H}_a H_b \sigma^{\mu\nu}F_{\mu\nu}Q^\xi_{ba}) +
 \frac{i g}{2}\, \mbox{Tr}\left(\bar{H}_aH_b\gamma_\mu\gamma_5(\xi^\dag\partial^\mu\xi - \xi\partial^\mu\xi^\dag)_{ba}\right),
\eeq
where Tr is a trace over the Dirac indices, and $\beta$ is a coupling constant parameterizing a local contribution to the
light quark magnetic moment. We include the most important one-loop correction, which is shown in Fig.~\ref{FIG:oneloop}.

The effective magnetic moment for the $B_s \rightarrow B_s^{*} \gamma$ transition is then 
\beq
\label{EQ:Bmueff}\mu_{B_s\gamma} = -\frac{1}{3m_b} - \frac{1}{3}\beta + g^2\frac{m_K}{4\pi f_K^2},
\eeq
where $g$ is the $\chi$PT coupling constant, and $m_K, f_K$ are the mass and decay constant of the Kaon respectively.
The constants $\beta$ and $g$ can be extracted from a combination of the experimental $D^{*+} $ branching ratios, 
$\mathcal{B}(D^{*+}\rightarrow D^+\gamma) = 0.016\pm0.004$ and $\mathcal{B}(D^{*+}\rightarrow D^0\pi^+) = 0.677\pm0.005$, 
and the total width, where the newest preliminary result from BaBar collaboration is reported to be 
$\Gamma_{D^{*+}} = 83.5\pm1.7\pm1.2$ KeV ~\cite{Godang:ICHEP2012}. The decay widths for these processes 
using the method above are given by 
\beqa
\Gamma(D^{*+}\rightarrow D^+\gamma) & \,=\, & \frac{\alpha_{EM}}{3}\left(\frac{2}{3m_c} - \frac{1}{3}\beta + 
g^2\frac{m_\pi}{4\pi f_\pi^2}\right)^2|\vec{k}|^3,\\
\Gamma(D^{*+}\rightarrow D^0\pi^+) & \,=\, & \frac{g^2}{6\pi f_\pi^2}|\vec{p_\pi}|^3.
\eeqa
This yeilds the approximate values of the coupling constants, $g \approx 0.552$ and $\beta \approx 7.29 GeV^{-1}$. 
With Eq.\ref{EQ:Bmueff} this gives us $|\mu_{eff}| \approx 1.13$ GeV$^{-1}$.

To complete evaluation of Fig.~\ref{FIG:pole} (a) in Eq.~(\ref{BstarContr}), we evaluate the $B_s^{*} \to \mu^+\mu^-$ transition.
The effective hamiltonian describing the weak $b\rightarrow s\ell^+\ell^-$ transition is 
\beqa\label{effham}
\nonumber\mathcal{H}_{b\rightarrow s \bar{\ell} \ell } =  \frac{G_F}{\sqrt{2}} V_{tb} V^{*}_{ts} \frac{e^2}{8 \pi^2} 
\biggl[\bar{s} \gamma^{\mu}(1& \,-\, &\gamma_5) b \cdot \bar{\ell}\left[ C^{eff}_{9V}(\mu,q^2) \gamma_\mu +C_{10A}(\mu^2) 
\gamma_\mu \gamma_5\right] \ell \\
\label{eq:VAcurrent1}& \,-\, & 2i m_b \frac{ C_{7\gamma}(\mu^2)}{q^2} q_\nu \cdot \bar{s} \sigma^{\mu\nu}(1+\gamma_5) b  
\cdot \bar{\ell}\gamma_\mu \ell\biggr],
\eeqa
where $q_\nu = (p_{\ell^+} + p_{\ell^-})_\nu$ is the momentum of the lepton pair and $C_i$ are scale-dependent Wilson
coefficients. The matrix element for $B_s^{*} \rightarrow \mu \bar{\mu}$ is then
\beqa
\mathcal{M}_{B_s^{*} \rightarrow \mu^+ \mu^-} &\,=\,& ~i\frac{G_F}{\sqrt{2}} V_{tb} V^{*}_{ts} \frac{e^2}{8 \pi^2} f_{B_s} M_{B_s} \biggl[\eta_{\mu}^{*} \bar{u}(p_{\mu^+})[C_9 \gamma^\mu +C_{10} \gamma^\mu \gamma_5] v(p_{\mu^-})  
\nonumber \\ 
&\,-\,& ~2 m_b \frac{C_7}{q^2}(\bar{u}(p_{\mu^+})\gamma_\mu v(p_{\mu^-}))q_\nu
(i\epsilon^{\mu\nu\alpha\beta}v_\alpha \eta_\beta + v^\mu \eta^\nu - v^\nu \eta^\mu)\biggr],
\eeqa
where $\eta^\mu$ and $v^\mu$ are the polarization and 4-velocity of the vector meson respectively. We defined
$\langle 0 | \bar{s}_L \gamma^{\mu} b_L | B_s^{*}\rangle = \eta^{\mu} f_{B_s^{*}}/2$, and 
$\langle 0 | \bar{s} \sigma^{\mu\nu}(1+\gamma_5) b | B_s^{*}\rangle = M_B 
f_{B_s}[i\epsilon^{\mu\nu\alpha\beta}v_\alpha \eta_\beta + v^\mu \eta^\nu - v^\nu \eta^\mu]$, 
with $f_{B_s^{*}} = M_{B_s} f_{B_s}$~\cite{Casalbuoni:1996pg}. This gives for the
amplitude of Fig.~\ref{FIG:pole} (a)
\beqa
\nonumber \mathcal{M}_{[B_s \rightarrow B_s^{*}\gamma_s \rightarrow \mu^+ \mu^- \gamma_s]} 
 &\,=\,& \frac{G_F}{\sqrt{2}} V_{tb} V^{*}_{ts} \frac{e^3}{8 \pi^2} \mu_{eff} \frac{f_{B_s} }{q^2 - M_{B_s^*}^2} 
 \left( \epsilon^{\mu \nu \alpha \beta} \epsilon_{\mu}^{*} k_{\alpha} q_{\beta}\right) \\
 &\,\times\,&\biggl[\left(2 C_7 m_b - C_9 M_{B_s^{*}}\right) [\bar{u}_{p_1} \gamma_{\nu} v_{p_2}] - 
C_{10} M_{B_s^{*}}  [\bar{u}_{p_1} \gamma_{\nu} \gamma_5 v_{p_2}]\biggr].
\eeqa
The other contribution that is leading the $M_{B_s} \to \infty$ limit is given in Fig.~\ref{FIG:pole} (c)
\beq
\mathcal{M}_{[B_s^{0} \rightarrow \mu \bar{\mu} \phi \rightarrow \mu \bar{\mu} \gamma_s]} = 
\mathcal{M}_{B_s^{0} \rightarrow \mu \bar{\mu} \phi} 
\times \frac{g_{\mu \nu}}{m_{\phi}^2} \times 
\mathcal{M}_{\phi \rightarrow \gamma_s},
\eeq

Employing vector-meson dominance, and using the definition of the vector meson decay constant 
$\langle 0 | \bar{s} \gamma^{\mu} s | \phi \rangle = f_{\phi} m_{\phi} \eta_{\phi}^{\mu}$, where $\eta_{\phi}^{\mu}$ is the 
polarization of the $\phi$ meson, and 
$\langle \gamma | \bar{s} (-i e Q_s \dslash A) s |\phi \rangle = (- i e Q_s ) \epsilon_{\mu}^{*} \langle 0 | \bar{s} \gamma^{\mu} s | \phi \rangle$,
\beq
\mathcal{M}^{\mu}_{\phi \rightarrow \gamma_s} = \frac{1}{3}\, e\, f_{\phi} \,m_{\phi} \, \epsilon_{\mu}^{*}.
\eeq
Again, we calculate $\mathcal{M}_{B_s^{0} \rightarrow \mu \bar{\mu} \phi}$ using (HM$\chi$pT). For the short distance 
contributions we use the effective Hamiltonian describing $b\rightarrow s \bar{\ell} \ell$ transitions in 
Eq.~\ref{effham}, as well as the effective Hamiltonian for $b \rightarrow s \gamma$,
\beqa\label{eq:tensorcurrent}
\mathcal{H}_{b \rightarrow s \gamma} \,=\,  \frac{G_F}{\sqrt{2}} 
V_{tb} V^{*}_{ts} \frac{e}{8 \pi^2} m_b C_{7\gamma}(\mu^2) \cdot \bar{s} \sigma^{\mu\nu}(1+\gamma_5) b \cdot F_{\mu \nu}.
\eeqa
In order to bosonize the quark currents found in Eqs.~(\ref{effham}) and (\ref{eq:tensorcurrent}) we introduce the light vector 
octet to the HM$\chi$pT~\cite{Casalbuoni:1996pg},
\beq
\rho_{\mu} \equiv i \frac{g_V}{\sqrt{2}} \left( \begin{array}{ccc}  \frac{\rho^0}{\sqrt{2}}+\frac{\omega}{\sqrt{2}} & \rho^+ & K^{*+} \\ \rho^- & -\frac{\rho^0}{\sqrt{2}}+\frac{\omega}{\sqrt{2}} & K^{*0}\\ K^{*-} & \bar{K}^{*0} & \phi \end{array}\right).
\eeq
The bosonized currents $\bar{s} \gamma^{\mu} (1-\gamma_5) b$ and $\bar{s} \sigma^{\mu \nu} (1+\gamma_5)b$ are,
respectively, 
\beqa
L^{\mu}_{1a} &=& \alpha_1 \langle \gamma_5 H_b (\rho^{\mu})_{bc} \xi_{ca}^{\dagger} \rangle, 
\nonumber \\
L^{\mu\nu}_{1a} &=& i \alpha_1 \biggl\{g^{\mu \alpha} g^{\nu \beta} - \frac{i}{2} \epsilon^{\mu \nu \alpha \beta}\biggr\} \langle \gamma_5 H_b [\gamma_{\alpha} (\rho_{\beta})_{bc} - \gamma_{\beta} (\rho_{\alpha})_{bc}] \xi_{ca}^{\dagger} \rangle.
\eeqa
A numerical value of $\alpha_1 = -0.07 \, GeV^{1/2}$~\cite{Casalbuoni:1996pg} will be used for our calculations. Keeping only the 
gauge invariant portion, the amplitude for the decay with an intermediate $\phi(1020)$ is
\beqa
\nonumber\mathcal{A}(B_s^{0} \rightarrow \mu \bar{\mu} \phi \rightarrow \mu \bar{\mu} \gamma_s) &\,=\,& G_F V_{tb} V_{ts}^{*} \frac{e^3 f_{\phi} g_{\phi} \alpha_1 C_7 m_b}{24 \pi^2 \sqrt{M_{B_s}} m_{\phi} (p_1 \cdot p_2)} \epsilon^{*}_{\mu} \biggl( (k \cdot (p_1 + p_2)) [\bar{u}_{p_1} \gamma^{\mu} v_{p_2}] \\
&\,-\,& (p_1 + p_2)^{\mu} [\bar{u}_{p_1} \not k v_{p_2}] + i \epsilon^{\mu \nu \alpha \beta}  k_{\alpha} (p_1 + p_2)_{\beta} [\bar{u}_{p_1} \gamma_{\nu} v_{p_2}]\biggr).
\eeqa
We checked that other contributions to the decay are smaller then the ones considered above.
We considered the bremsstrahlung diagrams where a soft photon is emitted from one of the outgoing leptons. 
These diagrams will result in an infrared divergence in the soft region, which has been shown to cancel with the 1-loop 
QED vertex corrections~\cite{Aliev:1997sd}. The vertex corrections, as well as the bremsstrahlung contributions, will remain 
suppressed by a power of the lepton mass. Therefore the remaining non-divergent contributions from both the 
bremsstrahlung and vertex corrections to final states with either an electron or a muon would not be significant.

The only contribution to the ampltitude from the effective hamiltonian describing the weak transition in Eq.(\ref{effham}) ends up being the 
$\mathcal{O}_{10}$ operator. This come from obtaining the matrix elements for the pseudoscalar meson,
 \beqa
 \langle 0 | (\bar{s}\gamma^{\mu} (1-\gamma_{5})b) | B_s\rangle  &\,=\,&  -i f_B P_B^{\mu},\\
\langle 0 | (\bar{s} \sigma^{\mu\nu}(1+\gamma_5) b) | B_s\rangle  &\,=\,&  0,
\eeqa
where $f_B$ is the decay constant of the B meson. With these definitions and using the conservation of the vector current we can arrive at an expression for the amplitude
 \beq
 \mathcal{M}_{Brem} = i e \frac{\alpha_{EM} G_F}{2\sqrt{2}} V_{tb}V_{ts}^{*} f_B C_{10} m_{\mu} \left[ \bar{\mu}\left( \frac{\dslash{\epsilon}\dslash{P_B}}{p_{\mu^-}\cdot k} - \frac{\dslash{P_B} \dslash{\epsilon}}{p_{\mu^+}\cdot k}\right) \gamma_{5} \mu \right],
\eeq
where $\epsilon^{\mu}$ and k are the polarization and momentum of the photon respectively. Just as we would expect from the helicity structure involved, the amplitude for the bremsstrahlung contribution is proportional to the lepton mass. So in the limit $m_\ell \rightarrow 0$, this contribution should be negligible compared to the non helicity-suppressed contributions.

Putting everything together, the distribution of the decay width as a function of the kinematic variable 
$s=(P_{B_s} - k)^2/M_{B_s}^2=q^2/M_{B_s}^2$, in the limit $m_{\ell} \rightarrow 0$,
\beqa\label{DifWidth}
\frac{d\Gamma}{ds} = \left. \frac{d\Gamma}{ds} \right |_{B_s^*} + 
 \left. \frac{d\Gamma}{ds}\right |_{\phi B_s^*}  + \left. \frac{d\Gamma}{ds}\right |_{\phi},
\eeqa
where the decay distributions are given for the two different resonance amplitudes and their interference.
\beqa\label{differentialwidth}
 \left. \frac{d\Gamma}{ds}\right |_{B_s^*}  &\,=\,&  ~ X_{CKM} M_{B_s}^3 f_{B_s}^2\mu_{eff}^2 
 \left[(|C_9|^2 +|C_{10}|^2)x_{B_s^{*}}+4 \, C_7^2 x_b - 4 \, C_7 C_9 x_b x_{B_s^{*}} \right]
 \frac{s (1-s)^3}{(s-x_{B_s^{*}}^2)^2},
\nonumber \\
 \left. \frac{d\Gamma}{ds} \right |_{\phi}  \phantom{q} &\,=\,& ~ X_{CKM} 
 \left[\frac{16\, C_7^2\, f_{\phi}^2\, g_{\phi}^2\, m_b^2\, \alpha_1^2}{9 \, m_{\phi}^2} \right] \frac{(1-s)^3}{s},
\\
 \left. \frac{d\Gamma}{ds}\right |_{\phi B_s^*}  &\,=\,& ~ X_{CKM}
\left[\frac{4 \sqrt{2} f_{B_s} f_{\phi} g_{\phi} M_{B_s}^{3/2} m_b \alpha_1 \mu_{eff}}{3 m_{\phi}} (C_7 C_9 x_{B_s^{*}}-2 C_7^2 x_b)
\right] \frac{(1-s)^3}{s- x_{B_s^{*}}^2},
 \nonumber
\eeqa
where we have defined $X_{CKM}=(G_F^2 |V_{tb} V^{*}_{ts}|^2 M_{B_s}^2  \alpha_{EM}^3)/(768 \pi^4)$,
$x_b \equiv m_b/M_{B_s}$, and $x_{B_s^{*}} \equiv M_{B_s^{*}}/M_{B_s}$. We use the Wilson coefficients 
$C_i(\lambda)$ choosing the scale at $\lambda \simeq m_b\simeq 5 GeV$, with $C_7 = 0.312$, $C_9 = -4.21$ 
and $C_{10} = 4.64$ \cite{Melikhov:2004mk}\cite{Grinstein:1988me}. The CKM matrix elements 
are $|V_{tb}V_{ts}^*| = (4.7\pm 0.8)\times 10^{-2}$ \cite{Beringer:1900zz}. With the most recent lattice calculation of 
$f_{B_s}$ is $\approx$ 228 MeV~\cite{Na:2012uh}. Note that, when integrated over the endpoint window the last two terms in Eq.~(\ref{DifWidth}) are much smaller than the first one. The interference contribution is destructive and is
\beqa
\mathcal{B}(B_s \rightarrow \mu^+ \mu^-\gamma_{E<60})_{\phi B_s^{*}} & \,=\, &  -5.0 \times 10^{-17},\\
\mathcal{B}(B_s \rightarrow \mu^+ \mu^-\gamma_{E<300})_{\phi B_s^{*}} & \,=\, & -1.1 \times 10^{-14},
\eeqa
which are both much smaller than the $B_s^{*}$ contribution alone. 

\begin{figure}
\begin{center}
{\includegraphics[width=12cm]{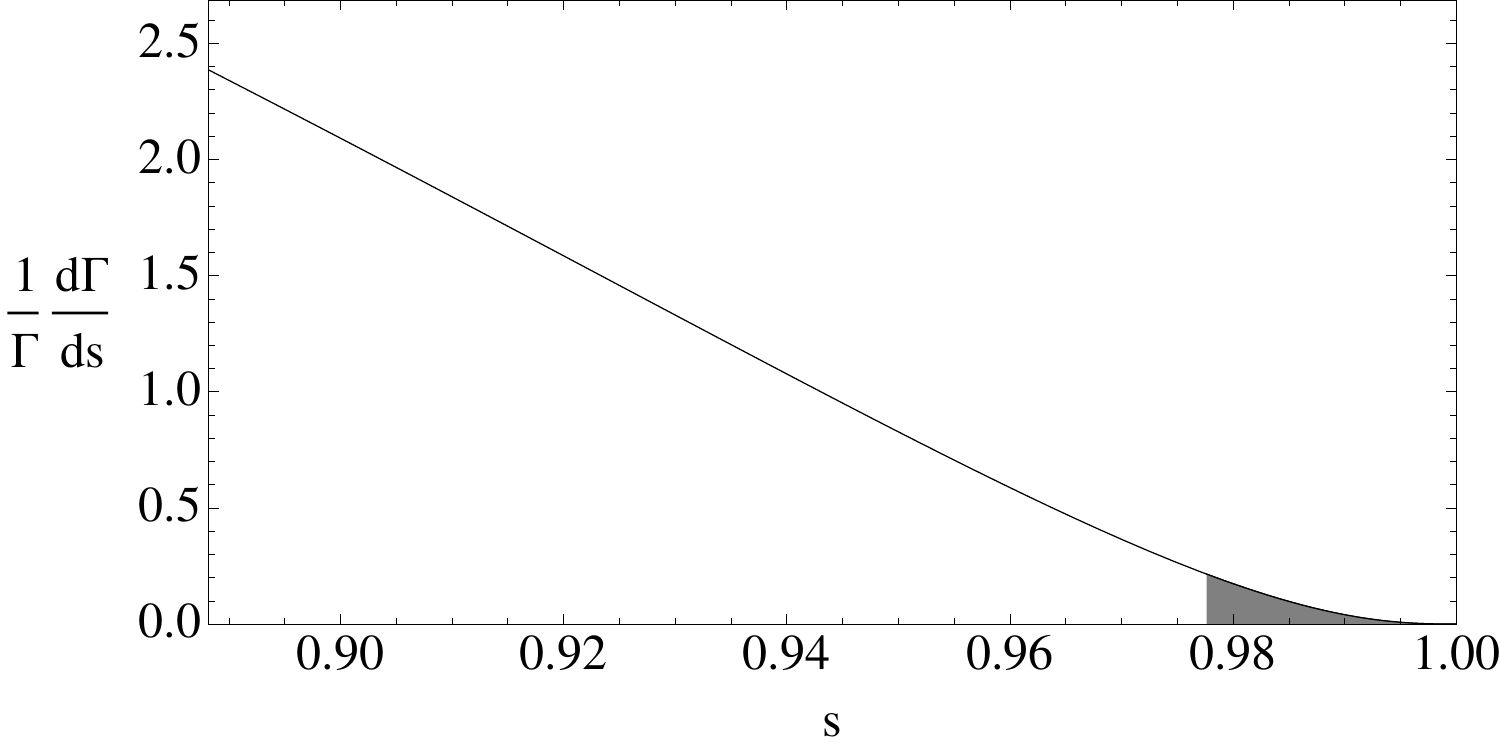}}
\end{center}
\caption{Normalized differential spectrum in s. The grey shaded region corresponds to the contribution from a soft photon energy cut at $E_\gamma \sim 60$ MeV.}
\label{FIG:dgds} % caption for the whole figure
\end{figure}

The normalized differential spectrum in $s$ is shown in Fig.(\ref{FIG:dgds}). The photon energy is related to the invariant mass as $E_\gamma = (1-s)M_B/2$, so we can integrate the differential spectrum over the required corresponding kinematic region in photon energy to obtain the decay width.

Integrating Eq.(\ref{DifWidth}) over the kinematic region corresponding to a soft photon cut of 
$E_\gamma \sim 60, 300$ MeV we get the respective branching ratios 
\beqa
\mathcal{B}(B_s \rightarrow \mu^+ \mu^-\gamma_{E<60}) & \,=\, & 1.6 \times 10^{-12},\\
\mathcal{B}(B_s \rightarrow \mu^+ \mu^-\gamma_{E<300}) & \,=\, & 1.1 \times 10^{-10},
\eeqa
which are quite too low to affect experimental determination of the branching ratio $B_s \to \mu^+\mu^-$, agreeing with the estimates of Ref. \cite{Buras:2012ru} where $\mathcal{B_{SM}}(B^0_s \rightarrow \mu^+ \mu^-) = (3.23\pm0.27)\times 10^{-9}$. 

%%%%%%%%%%%%%%%%%%%%%%%%%%%%%%%%%%%
\section{$B^0_s \to \mu^+\mu^-\nu_\mu\bar \nu_\mu$ transition}

\begin{figure}
\begin{center}
\includegraphics[width=14cm]{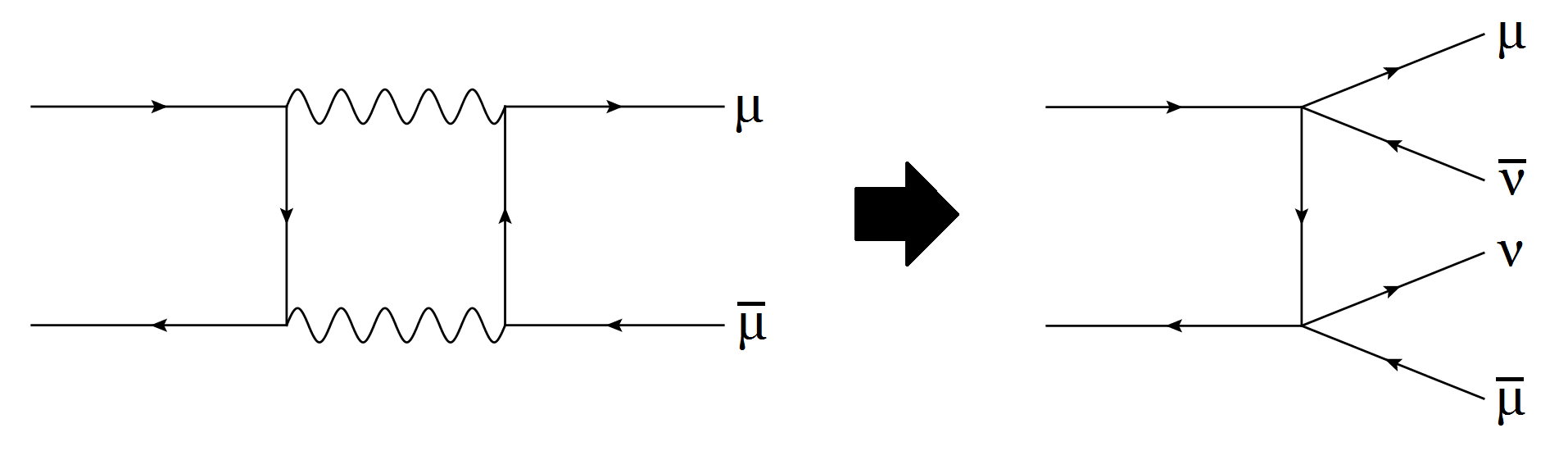}
\end{center}
\caption{$B_s \rightarrow \mu^+ \mu^- \bar{\nu} \nu$}
\label{FIG:brokenbox} % caption for the whole figure
\end{figure}

Because of the Glashow-Illiopulous-Maiani (GIM) mechanism, the SM loop diagram for the helicity-suppressed 
$B_s^0\to \mu^+\mu^-$ decay is dominated by the intermediate top quark despite being suppressed by the 
CKM factors $|V_{tb} V^{*}_{ts}|^2$. A transition similar to the ones described above, which on a portion of the 
available phase space looks like $B_s^0\to \mu^+\mu^-$ is the tree-level decay 
$B^0_s \rightarrow \mu^{+} \mu^{-}\nu\bar{\nu}$. The dominant tree-level contribution for this process is 
depicted in Fig.~ \ref{FIG:brokenbox}. This decay can have a contribution to the background, which appears only below 
$q^2 = M_{B_s}^2$ and, if numerically significant, can affect the extraction of  $\mathcal{B}(B_s \rightarrow \mu^+ \mu^-)$.
This process is neither loop-dominated nor is it helicity suppressed.  It nevertheless has a kinematic phase space 
suppression due to the four-particle final state. For the $B_s$ meson decay, an intermediate charm quark will give the 
largest contribution since the intermediate top quark diagram will be suppressed by the mass of the top quark. 
Also, the up quark contribution is suppressed by $V_{ub} V_{us}^{*} \approx \lambda^4$ whereas the charm contribution 
is only suppressed by $V_{cb} V_{cs}^{*} \approx \lambda^2$, where $\lambda \approx 0.22$. 

The transition amplitude for this process is simple, 
\beqa
\mathcal{M}_{B_s \to \mu^+\mu^- \nu\bar\nu} =  \frac{G_F^2}{2} V_{cb} V^{*}_{cs} \langle 0 | \bar{s} \gamma_{\beta} (1-\gamma_5)\frac{ i (\dslash{p_c} + m_c)}{p_c^2-m_c^2} \gamma_{\alpha}(1-\gamma_5) b | B_s \rangle L^{\alpha}_1 L^{\beta}_2,
\eeqa
where $L^{\alpha} = \bar{\mu} \gamma^{\alpha} (1-\gamma_5) \nu_\mu$. In the rest frame of the decaying meson we 
can reduce the phase space integral's dependence to five independent  Lorentz invariants.  In the same fashion 
as in \cite{Axelrod:1983xs} we define these invariants as
\beqa
\nonumber S_{12} = (p_{\mu^-}+p_{\mu^+})^2, \quad S_{13} & \,=\, & (p_{\mu^-}+p_{\bar{\nu}})^2, \quad S_{34} = (p_{\bar{\nu}}+p_{\nu})^2 , \\
  S_{123} = (p_{\mu^-}+p_{\mu^+}+p_{\bar{\nu}})^2& \,,\, & \quad S_{134} = (p_{\mu^-}+p_{\bar{\nu}}+p_{\nu})^2.
\eeqa
Our width then becomes
\beqa
d\Gamma = \frac{(2 \pi)^4}{2 M} \int \left(\frac{\pi^2}{2 M^2}\right)
 \frac{\left|\mathcal{M}_{B_s \to \mu^+\mu^- \nu\bar\nu}\right|^2}
 {\left[-\Delta_4(p_{\mu^-}, p_{\mu^+}, p_{\bar{\nu}}, p_{\nu})\right]^{1/2}} 
 dS_{12} dS_{123} dS_{13} dS_{134}, 
\eeqa
where $\Delta_4$ is the symmetric Gram determinant
\beqa
\Delta_4(q,r,s,t) =  \left| \begin{array}{cccc}
q^2 & \quad q \cdot r & \quad q \cdot s & \quad q \cdot t \\
r \cdot q & \quad r^2 & \quad r \cdot s & \quad r \cdot t\\
s \cdot q & \quad s \cdot r & \quad s^2 & \quad s \cdot t\\
t \cdot q & \quad t \cdot r & \quad t \cdot s & \quad t^2\end{array} \right|.
\eeqa
In order to avoid the divergence of $1/(-\Delta_4)^{1/2}$ on the boundary, suitable variable changes can be made 
thereby making the singularity integrable. We define
\beqa
\nonumber S_{134} & \,=\, & \frac{1}{2a} \left[-b+\sin (\tilde{S}_{134})(b^2-4 a c)^{1/2}\right],\\
S_{13} & \,=\, & 4 (-a)^{1/2} \tilde{S}_{13} + m_{\ell}^2,
\eeqa
where $a, b$ and $c$ are the parameters solved by 
\beqa
-\Delta_4(p_{\mu^{-}}, p_{\mu^{+}}, p_{\bar{\nu}}, p_{\nu}) = a S_{134}^2 + b S_{134} + c\,.
\eeqa
The limits of integration are calculated in \cite{Axelrod:1983xs}, resulting in our partial width
\beqa
\frac{d\Gamma}{dS_{12}} = \frac{2}{(4 \pi)^6 M^3} \int_{S_{12}}^{M^2} dS_{123} 
\int_{0}^{\xi} dS_{34} \int_{m_{\ell}^2/S_{12}}^{1} 
d\tilde{S}_{13} \int_{-\pi/2}^{\pi/2} d\tilde{S}_{134} \left|\mathcal{M}_{B_s \to \mu^+\mu^- \nu\bar\nu}\right|^2,
\eeqa
where $\xi = (M^2-S_{123})(S_{123}-S_{12})/S_{123}$.
We define the cut on missing energy as $S_{12}^{cut}(E_{cut}) = M^2- 2 M (E_{cut})$ which gives us a lower limit on 
$S_{12}$ for the final integral in order to obtain the decay width. The branching ratios for this contribution can then 
calculated using numerical phase-space integration for various cuts including the one that corresponds to the 
invariant mass range seen at the LHCb \cite{:2012ct}.
\beqa
\nonumber BR\left[B_s \rightarrow \mu^{+} \mu^{-} \nu \bar{\nu}\right]_{E_{cut} = 60 MeV} & \,=\, & 1.6 \times 10^{-25}\\
BR\left[B_s \rightarrow \mu^{+} \mu^{-} \nu \bar{\nu}\right]_{E_{cut} = 300 MeV} & \,=\, & 1.4 \times 10^{-18}.
\eeqa
As we can see, the due to enormous phase space suppression (we are only interested in a small sliver of the 
available four-particle final state), the possible contribution from this decay is unimportant for experimental analyses.

%%%%%%%%%%%%%%%%%%%%%%%%%
\section{Conclusions}\label{Conclusions}

Recent evidence of observation of the flavor-changing neutral current decay $B^0_s \rightarrow \mu^+ \mu^-$ by
the LHCb collaboration $\mathcal{B}(B^0_s \rightarrow \mu^+ \mu^-) = (3.2^{+1.5}_{-1.2})\times 10^{-9}$ is
consistent with the latest standard model predictions. We analyzed branching ratios of the decays 
$B^0_s \rightarrow \mu^+ \mu^- X$, with $X = \gamma$ or $\nu\bar\nu$, which can mimic 
$B^0_s \rightarrow \mu^+ \mu^-$ on portions of the parameter space where $X$ is soft. We performed a 
model-independent calculation of those processes incorporating heavy quark and chiral symmetries of QCD. 
Our calculations concentrated on the contributions that are not helicity suppressed by powers of $m_{\mu}$ and leads to a correction to the SM prediction of approximately 3$\%$ at a photon energy cut of 300 MeV and less than 1$\%$ at a cut of 60 MeV from soft photon contributions to the decay $B_s \rightarrow \mu^{+} \mu^{-}$.  
The possible contamination from $B_s \rightarrow \mu^{+} \mu^{-} \nu \bar{\nu}$ is even smaller, at the 
sub percent level.

We would like to thank Gil Paz and Rob Harr for useful discussions. This work was supported in part by the 
U.S.\ Department of Energy under Contract DE-SC0007983.

%%%%%%%%%%%%%%%%%%%%%%%%%%%%%%%%%%%

%\pagebreak

%%%%%%%%%%%%%%%%%%%%%%%%%%%%%%%%%%%%%%%%%%%%

\end{document}